\documentclass[aps,superscriptaddress,twocolumn,floats,tightenlines,balancelastpage,floatfix]{revtex4}
\input epsf
\usepackage{amsmath,amssymb}
\usepackage{graphicx}

\newcommand{\beq}{\begin{equation}}
\newcommand{\eeq}{\end{equation}}
\newcommand{\beqa}{\begin{eqnarray}}
\newcommand{\eeqa}{\end{eqnarray}}

\begin{document}

\title{Anisotropy of the Cosmic Neutrino Background}
\author{R.~J. Michney and R.~R. Caldwell \\Department of Physics \& Astronomy, Dartmouth College, 
Hanover, NH 03755}

%\author{R.~J. Michney} \author{R.~R. Caldwell} 
%\address{Department of Physics \& Astronomy, Dartmouth College, Hanover, NH 03755}

\date{\today}

%%%%%%%%%%%%%%%%%%%%%%%%%%%%%%%%%%%%%%%%%%%%%%%%%%%%%%%%%%%%%%%%%%%%%%%%%%%%%%%%%%%%%%%%%%%%%

\begin{abstract} 
The cosmic neutrino background (CNB) consists of low-energy relic neutrinos
which decoupled from the cosmological fluid at a redshift $z \sim 10^{10}$.
Despite being the second-most abundant particles in the universe, direct
observation remains a distant challenge. Based on the measured neutrino mass
differences, one species of neutrinos may still be relativistic with a thermal
distribution characterized by the temperature $T\sim 1.9$K. We show that the
temperature distribution on the sky is anisotropic, much like the photon
background, experiencing Sachs-Wolfe and integrated Sachs-Wolfe effects.
\end{abstract}

\maketitle

%%%%%%%%%%%%%%%%%%%%%%%%%%%%%%%%%%%%%%%%%%%%%%%%%%%%%%%%%%%%%%%%%%%%%%%%%%%%%%%%%%%%%%%%%%%%% 
  
Relic particles from the early universe carry a wealth of information about the origin and history of
the cosmos. Relic photons, in the form of the cosmic microwave background (CMB), have revealed the
conditions in the universe back to a redshift $z\sim 1100$, when the universe was $\sim 400,000$
years old. Relic gravitational waves from inflation are anticipated to provide a snapshot of the
cosmos at $z\sim 10^{27}$, just $\sim 10^{-35}$ seconds after the Big Bang. In the present
investigation, we consider the cosmic neutrino background (CNB), a sea of relic neutrinos which carry
information about the conditions in the universe at a redshift $z\sim 10^{10}$ and time $t\sim 1$
sec. We expect the CNB to be characterized by a Fermi-Dirac distribution at temperature $T =1.9$~K,
with slight anisotropies owing to inhomogeneities in the universe at, and since, $z\sim 10^{10}$. Our
goal is to calculate the CNB anisotropy spectrum.  

The CNB was formed when the neutrino sea dropped out of thermal equilibrium with the other
matter and radiation of the early universe. The Standard Model neutrinos, $\nu_i$ for
$i=\{e,\,\mu,\,\tau\}$, were coupled to the electron content of the early universe 
primarily through the interaction $\nu_i+\bar{\nu}_i\leftrightarrow e^- +e^+$, for which the
neutrino-antineutrino cross section is the limiting factor. The average annihilation rate
\beqa
\Gamma &=& \frac{16 \,G_F^2}{\pi^3} (g_L^2 + g_R^2) T^5 \\
g_L^2 + g_R^2 &=& 
\begin{cases}
\sin^4 \theta_W + (\frac{1}{2}+\sin^2\theta_W)^2  
&\text{for $\nu_e$} 
\cr 
\sin^4 \theta_W +(-\frac{1}{2}+\sin^2 \theta_W)^2  
&\text{for $\nu_{\mu,\tau}$}
\end{cases} \nonumber
\eeqa
kept the neutrinos in good thermal contact with the cosmic fluid until $\Gamma\sim H$. (See
Refs.~\cite{KolbTurner,Hannestad:2006zg}.) Since the universe expands with temperature as
\beq
H(T)=\frac{1.66g_*^{1/2} T^2}{M_{Pl}}\,,
\eeq
we obtain decoupling temperatures of $T_{\nu_e}=2.4$~MeV and $T_{\nu_{\mu,\tau}}=3.7$~MeV,
using accepted Standard Model parameters. These temperatures correspond to a redshift $z\sim
10^{10}$, occuring very shortly before $e^-e^+$ freeze-out, at $T\simeq m_e\,/\, 3$.
Thereafter, the CNB evolved as a Fermi-Dirac distribution with a cooling temperature  $T_\nu
= (4/11)^{1/3} T_\gamma$ relative to photons. The present-day value is  $T_\nu = 1.946$~K. 

All three species of CNB neutrinos remained relativistic until the temperature dropped below
their rest mass. Cosmological bounds indicate the mass of the heaviest neutrino to be
$0.04\,{\rm eV} \lesssim m_{\nu_i} \lesssim (0.2 - 0.4)\, {\rm eV} $ \cite{Kayser} while
measurements of neutrino mass differences yield $\Delta m_{12}^2 \approx 8\times
10^{-5}\,{\rm eV}^2$ (solar neutrinos) and $\Delta m_{23}^2 \approx 2.5 \times 10^{-3}\,{\rm
eV}^2$ (atmospheric neutrinos), establishing that at least two neutrino flavors have masses
$\gtrsim 10^{-2}$~eV \cite{SKref,SNOref,Kamref}. These results allow for the possibility
that one mass eigenstate has $m < T_\nu = 1.6\times 10^{-4}$~eV and therefore remains
relativistic. For this investigation we assume one surviving relativistic species.

The predicted variations in the CNB intensity are obtained from its phase-space distribution,
\begin{equation}
f(x^i,\,P_j,\,\tau) = f_0(q)[1 + \Psi(x^i,\,q,\,n_j,\,\tau)],
\end{equation}
where $f_0(q)$ is the background, Fermi-Dirac neutrino distribution at momentum $q$. The
neutrino temperature perturbation is $\Delta = -\Psi (d\ln f_0/d\ln \epsilon)^{-1}$. We
assume that neutrino decoupling takes place instantaneously. Hence, the perturbation $\Psi$,
at comoving location $x^i$ and conformal time $\tau$ for neutrinos moving in the direction
$n_j$, evolves according to the collisionless Boltzmann equation, 
\begin{equation}
\frac{\partial \Psi}{\partial\tau} + i \frac{q}{\epsilon}(\vec k \cdot \hat n) \Psi
+ \frac{d\ln f_0}{d\ln q} \left[ \dot\phi - i \frac{\epsilon}{q}
(\vec k \cdot \hat n)\psi\right] = 0.
\end{equation}
We follow the notation of Ref.~\cite{Ma:1995ey} where $\epsilon = \sqrt{q^2 + a^2 m^2}$, $a$
is the expansion scale factor normalized to unity at present, and $\phi,\,\psi$ are the
gravitational potentials in the conformal-Newtonian gauge. Hereafter we assume that
anisotropic stress perturbations are negligible, so that $\psi=\phi$. Parametrizing the
neutrino's flight with conformal variable $\lambda$, we can write  $-\partial_\lambda =
\partial_\tau + \frac{q}{\epsilon}\hat n \cdot \vec\nabla$ for the derivative along the path
of a neutrino from decoupling to the observer. Then defining $\Gamma \equiv d\ln f_0/d\ln
\epsilon$, the Boltzmann equation simplifies to
\beq
\partial_\lambda \left(\Gamma \Delta\right) + \Gamma\left[ \partial_\lambda + \left(
\frac{q^2}{\epsilon^2} + 1\right) \partial_\tau\right] \phi = 0.
\eeq
Integrating along the line-of-sight $\lambda$ from decoupling to the present, we find the solution
\beqa
\Delta_0 &=& -\phi_0 + \frac{\Gamma_{dec}}{\Gamma_0}\left(\Delta_{dec} + \phi_{dec}\right)\cr
&+& \frac{1}{\Gamma_{0}}\int_{dec}^{0}d\lambda\, 
\left[\Gamma_{,\lambda}\,\phi - \Gamma (\frac{q^2}{\epsilon^2} + 1)\phi_{,\tau}\right].
\label{inteqn}
\eeqa
We may neglect $\phi_0$, which contributes only to the temperature anisotropy monopole. The
remaining terms on the first line give the anisotropy due to the initial temperature
fluctuations and gravitational potential at decoupling. In the limit of relativistic
particles, for which $\Gamma_{dec}=\Gamma_{0}=$~constant, this last term corresponds to  the
Sachs-Wolfe effect (SW) \cite{SachsWolfe}. The terms on the second line give the anisotropy
due to line-of-sight variations in the spectral shape, $\Gamma$, and gravitational
potential. In the relativistic limit only the latter term survives, in the form of the
integrated Sachs-Wolfe effect (ISW) \cite{Rees}. 

Decoupling occurs deep in the radiation era when the neutrinos are relativistic  and the
neutrino density perturbation contrast mirrors the total energy density perturbation
contrast, $\delta_\nu = \delta$.  In turn, the total density contrast is proportional to the
gravitational potential, $\delta = -2\phi$, which is a constant on large scales.
Consequently, the initial perturbations can be expressed in terms of the gravitational
potential at decoupling,
\beq
\Delta + \phi\,|_{dec} = \frac{1}{4}\delta_\nu + \phi\,|_{dec} = \frac{1}{2}\phi|\,_{dec}\,.
\eeq
Applying this result to (\ref{inteqn}), we find that the present-day, large-angle
temperature anisotropy is
\beq
\Delta_0 = \frac{1}{2}\frac{\Gamma_{dec}}{\Gamma_0} \phi_{dec} +
\frac{1}{\Gamma_{0}}\int_{dec}^{0}d\lambda\, 
\left[\Gamma_{,\lambda}\,\phi - \Gamma (\frac{q^2}{\epsilon^2} + 1)\phi_{,\tau}\right].
\label{fulleqn}
\eeq
For relativistic neutrinos, the large-angle temperature anisotropy is due to SW and ISW
contributions, $\Delta_0 = \frac{1}{2}\phi_{dec}-2 \int_{dec}^0 d\lambda\,\phi_{,\tau}$.

Let us consider the anisotropy arising solely from the gravitational potential at
decoupling. In this case, the temperature pattern in a direction $\hat n$ on the sky is
given by
\beq
\Delta_0(\hat n) = \frac{1}{2} \frac{\Gamma_{dec}}{\Gamma_0} \phi_{dec}(\hat n)\, .
\eeq
The ratio of spectral shape functions is
\beq
\frac{\Gamma_{dec}}{\Gamma_0}  \approx \frac{q}{\sqrt{q^2 + m^2}}
\frac{1 - e^{-\sqrt{q^2 + m^2}/kT_0}}{1 - e^{-q/kT_0}}
\eeq
where we take $m\, a_{dec} \ll q$. In the case of neutrinos which are non-relativistic
today, $q \ll m$ so that $\Gamma_{dec} \ll \Gamma_0$. Hence, the temperature anisotropy is
suppressed by a factor $\sim(q/m)^2$. However, in the case of relativistic neutrinos, 
$\Gamma_{dec} = \Gamma_0$, so this contribution to the temperature anisotropy is similar to
the CMB Sach-Wolfe effect, but with two notable differences. First, the CNB has a  prefactor
to the gravitational potential correlation of $1/4$ where the CMB has $1/9$, reflecting the
difference in the equation of state of the dominant form of energy (matter in the case of
the CMB, and radiation in that of the CNB) at the time the background is emitted. Second,
although the long-wavelength gravitational potential is a constant in both the radiation and
matter eras, the constant differs by a factor of $9/10$ as the potential decays by $10\%$
across the radiation-matter transition. (For simplicity, we ignore the effect of neutrino
anisotropic stress on the evolution of perturbations.) Including the difference in the mean
temperature of the background, 
\beq
\Delta T\,|_{CNB,SW} = \frac{3}{2}\times\frac{10}{9}
\times \left(\frac{4}{11}\right)^{1/3} \times \Delta T\, |_{CMB,SW}\,,
\eeq
we see that the SW temperature anisotropy in the CNB is $\sim 1.2$ times as strong as in the
CMB.

Next consider the anisotropy arising along the neutrino path. There are two such terms in
equation (\ref{inteqn}), one arising from $\Gamma_{,\lambda}$ and the second from
$\phi_{,\tau}$. To estimate the magnitude of the first term, we note that $\Gamma$ is nearly
a constant while the neutrinos are still relativistic, so the contribution at early times is
negligible. At late times, $\Gamma_{,\lambda}\propto {\cal H} \Gamma_0$ where the constant
of proportionality is of order unity for neutrinos with $q\simeq k T_0$. The resulting
contribution to the temperature anisotropy is $\Delta_0 \propto\, \int^0 d\lambda\, \phi
{\cal H}$, very similar to the standard ISW. Considering the second term, the only
significant contribution to the temperature anisotropy occurs at late times, when
$\Gamma/\Gamma_0 \to 1$  and the gravitational potential evolves due to the onset of
accelerated cosmic expansion. At these late times, $q\ll\epsilon$ so the nonrelativistic ISW
is approximately half the amplitude of the standard ISW.

In the case of relativistic neutrinos, the ISW effect is nearly the same as for photons,
with one difference. Neutrino decoupling takes place in the radiation era, so neutrinos
receive an additional ISW contribution due to the time-varying potential across the
radiation-matter transition. The photons do not fully experience this early-ISW effect
because CMB last scattering takes place at the tail end of this transition. Considering
only the late-time ISW, after $z \lesssim 10$,
\beq
\Delta T\,|_{CNB,ISW} = \left(\frac{4}{11}\right)^{1/3} \times \Delta T\, |_{CMB,ISW}
\eeq
so the CNB ISW is smaller by a factor of $\sim 0.7$ than the CMB. However, these rough estimates
ignore the early ISW effect for the CNB, the interference between the SW and ISW
contributions, and the wavelength dependence of the gravitational potential.

\begin{figure}[t]
\includegraphics[width=3.25in]{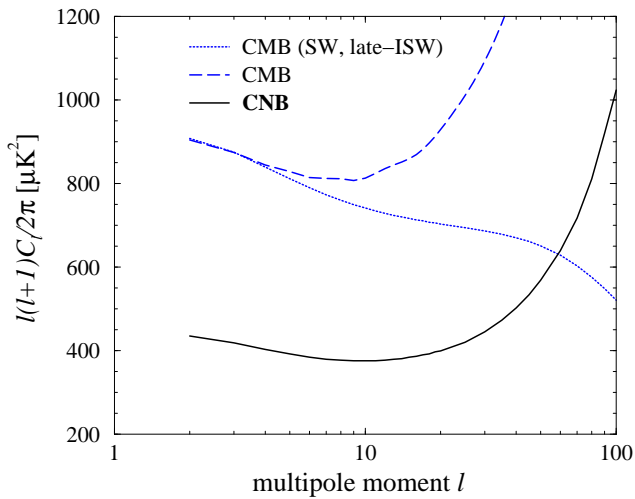}
\caption{The temperature anisotropy power spectrum for a species of relativistic neutrinos
is shown for multipole moments $\ell < 100$. The sources of anisotropy are SW and ISW
effects, including intrinsic perturbations of the neutrino spectrum at decoupling, which
occured at $z\sim 10^{10}$. For comparison, we also show the SW and late-ISW contributions to the
CMB anisotropy spectrum calculated following the same method. The amplitude is set by using $A_{WMAP}=0.9$
\protect{\cite{Spergel:2006hy,CMBfast}}. We also show the full CMB power spectrum calculated
using CMBfast \protect{\cite{Seljak:1996is}}.}
\label{fig1}
\end{figure}
 
We now present the results of detailed calculations of the CNB temperature anisotropy
spectrum. We assume relativistic neutrinos, so that the temperature fluctuations are given
by
\beqa
\Delta_0(\hat n) &=& \frac{1}{2}\phi(\tau_{dec},(\tau_0 - \tau_{dec})\hat n)  \cr
&+&2\int^{\tau_0 -\tau_{dec}}_{0}d\lambda\,  \phi_{,\tau}(\tau_0-\lambda,\lambda\hat n).
\eeqa
We assume a primordial spectrum of scale-invariant density perturbations, where the
Fourier modes of the gravitational potential obey $\langle \phi(\vec k)\phi(\vec k')\rangle
= 4\pi k^3 P_\phi(k)\delta(\vec k + \vec k')$, with a power spectrum $P_\phi(k) = A
k^{n_s-4}$. For the power-law index, $n_s=1$.
The background cosmology is modeled as a spatially-flat FRW spacetime filled by a
three-component fluid, consisting of radiation, matter, and cosmological constant. The
Hubble constant evolves as
\beq
H(a) = H_0\sqrt{\Omega_m ({a_0}/{a})^3 + \Omega_r ({a_0}/{a})^4 + \Omega_\Lambda}
\eeq
where we use $H_0 = 100\,h$~km/s/Mpc, $h=0.7$, $\Omega_m=0.3$, $\Omega_r=4.2\times
10^{-5}/h^2$, and $\Omega_\Lambda=1-\Omega_m - \Omega_r$. The evolution of the gravitational
potential due to adiabatic density perturbations is determined by \cite{Mukhanov,Ma:1995ey}
\beq
\ddot\phi + 3 {\cal H}(1 + c_s^2)\dot\phi + \left(2 \dot{\cal H} + (1 + 3 c_s^2) {\cal H}^2
+ c_s^2 k^2 \right)\phi = 0
\eeq
where the dot indicates the derivative with respect to conformal time, ${\cal H} = \dot a/a = a H$,
and the adiabatic sound speed is $c_s^2 = \frac{1}{3}/(1 +
\frac{3}{4}\frac{\rho_m}{\rho_r})$. Initial conditions are chosen so that
$\phi=1,\,\dot\phi=0$ deep in the radiation era. Last scattering occurs sharply at
$z=10^{10}$ for the CNB ($1100$ for the CMB). Finally, the $C_\ell$'s are obtained by
evaluating
\begin{widetext}
\begin{equation}
C_\ell = (4 \pi)^2 A \, 
\int d\ln k \, \left(\frac{1}{2}\phi(\tau_{dec},k) j_\ell(k (\tau_0 -\tau_{dec})) 
+ 2 \int_0^{\tau_0 - \tau_{dec}} d\lambda \, \phi_{,\tau}(\tau_0-\lambda,\lambda) 
j_\ell(k\lambda)\right)^2.
\end{equation}
\end{widetext}
To normalize to WMAP, we set the constant $A=200 \pi A_{WMAP}({T_\nu}/{T_{CMB}})^2 \mu K^2$.
The resulting multipole spectrum is shown in Figure~\ref{fig1}, our main result. It is
interesting to note that, whereas the CNB SW effect is much stronger than for the CMB, a
strong, negative cross-correlation between the early-ISW and the SW effects greatly reduces
the overall anisotropy power spectrum on large angular scales. We limit our
power spectrum to $\ell < 100$; on smaller angular scales we expect the effects 
of bulk anisotropic pressure or shear, which we have ignored, to be important.
For comparison, we also show the SW and late-ISW contributions to the CMB, as well as the full CMB. 
%XXX
These results are consistent with the earlier results of Hu {\it et al.} \cite{Hu:1995fq}, where
the CNB anisotropy in a SCDM universe was considered.
%XXX
Overall, the rms
temperature fluctuations in the CNB are smaller than for the CMB.

\begin{figure}
\includegraphics[width=3.25in]{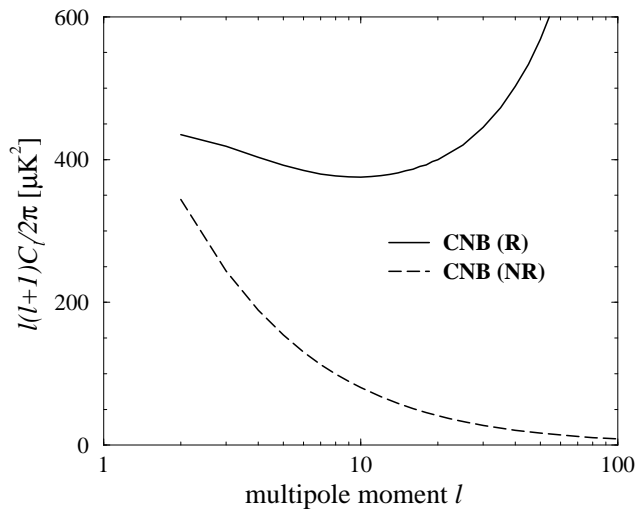}
\caption{The temperature anisotropy power spectrum for a species of non-relativistic (NR)
neutrinos is shown. We use neutrino mass $m_\nu = 0.05$~eV and momentum $q=T_\nu$. The
results do not change appreciably for $0.1\le q/T \le 10$. The SW effect is strongly
suppressed, and only a line-of-sight term contributes to the anisotropy. For comparison, we
also show the relativistic (R) CNB anisotropy spectrum.}
\label{fig2}
\end{figure}

The anisotropy power spectrum for non-relativisitic neutrinos is shown in Figure~\ref{fig2}. Since
the spectrum depends on the momentum of the neutrinos, we focus on values of momentum $q \sim T_\nu$
for comparison with anisotropies at the peak of the relativistic neutrino flux spectrum. We see that
the overall amplitude is comparable to the relativistic result at the lowest multipoles, as seen in
Figure~\ref{fig2}, but is much smaller otherwise, due to the suppression of the SW contribution from
decoupling. The dominant effect is due to the $\Gamma_{,\lambda}\phi$ term in the line-of-sight
integration in equation (\ref{fulleqn}), whereas the $\phi_{,\tau}$ term contributes a much smaller
fraction than in the relativistic case. In the limit $q \gg m_\nu > T_\nu$ the relativistic result is
obtained. Although low energy neutrino capture by galaxies is an anisotropic effect which removes the
slowest-moving particles from the power spectrum, the mass estimates used and the currently favored
upper bounds predict a negligible amount of gravitational clustering.

Our analysis thus far treats the primordial neutrinos analogously to CMB photons. However, there are several
important ways in which neutrinos differ from photons: exclusive interaction via the weak force,
spin-statistics, mass, and flavor oscillation. The first factor has been taken into account in determining the
neutrino decoupling time, and ignoring important CMB phenomena such as reionization. Even in the early
universe, when conditions were extremely hot and dense, neutrino scattering was negligibly rare. Present day
interactions with interstellar matter are far less significant in comparison with electromagnetic processes.
The CNB should also differ from the CMB in the absence of an acoustic peak. The second factor means that
neutrinos obey Fermi-Dirac statistics. The primordial, relativistic neutrinos have a thermal, blackbody
spectrum, but very low momentum neutrinos could be degenerate, a possibility which we will not explore
\cite{DegenerateNeutrinos}. The third factor is significant, as only one of the neutrino mass states may be
relativistic today.  This also leads to the fourth, final factor: flavor oscillation. The neutrino flavor
states $\nu_{j}$, $j=\{e,\,\mu,\,\tau\}$ are a superposition of mass eigenstates, $\nu_{i}$, $i=\{1,\,2,\,3\}$.
Hence, the flavor of an individual neutrino oscillates with time.

The CNB neutrinos have travelled over $\sim 45$ Gpc from the time of decoupling to the
present. Very early on, the universe was dense enough to be characterized as
electron-flavored matter, a distinction which determined the type of matter oscillations.
However the vast majority of the subsequent travel by neutrinos has occurred in vacuum,
where the path-length for neutrino mixing is very short. We can use the two-flavor mixing
model as an approximation, whereby the mixing probability is
\beq
P(\nu_i\to\nu_j) = \sin^2 2\theta_{ij} \sin^2\left(\frac{\Delta m_{ij}^2 L}{4 E_\nu} \right)
\eeq
Using $E_\nu \sim T_\nu$ we find a microscopic oscillation length. Hence, the CNB should be
well mixed \cite{Khlopov1981}, with flavor abundances dictated by the mixing angles $\theta_{ij}$. This means
the spectrum of primordial $\nu_e$'s should be a mixture of relativistic $\nu_1$'s and
nonrelativistic $\nu_{2,3}$'s. Similarly, the anisotropy spectrum should be a mixture. 
  
The anisotropy power spectrum for the $e,\,\mu,\,\tau$-neutrinos will be a mixture of the
spectra for the relativistic and nonrelativistic neutrino mass eigenstates. Since the
anisotropy for the relativistic species rapidly dominates with increasing multipole moment, 
we can ignore the nonrelativistic
contributions to leading order. Also, sufficiently low-momentum, nonrelativistic neutrinos
will get gravitationally captured in galaxies and clusters. Since we are assuming an
inverted neutrino mass spectrum, the lightest neutrino is $\nu_1$. We refer to the lepton
mass mixing matrix to determine what fraction $\nu_1$ contributes to the flavor eigenstates,
and to the temperature anisotropy in a particular flavor. For $\nu_e$, this matrix element
is $c_{12} c_{13}$ where $c_{ij} \equiv \cos\theta_{ij}$, so the temperature anisotropy of
electron neutrinos is $\Delta T(\nu_e) = c_{12} c_{13} \Delta T(\nu_1)$. The power spectrum
is quadratic in this fraction, $C_l(\nu_e) = (c_{12} c_{13})^2 C_l(\nu_1)$. Using currently
accepted values of the mixing angles, $\theta_{12} \approx 34^\circ$ and $c_{13} \approx 1$
\cite{Kayser}, then the electron neutrino temperature anisotropy power spectrum is
approximately $70\%$ of the relativistic CNB amplitude given in Figure~\ref{fig1}.

Thus, we have determined the large-angle CNB temperature anisotropy. It is important to note that
there is an extensive literature on cosmological neutrinos (e.g. \cite{Lesgourgues:2006nd}) including
their effect on the CMB. In fact,the existence of the CNB \cite{Dicke1965,Gerstein1966} was
immediately anticipated following the discovery of the CMB by Penzias and Wilson \cite{Penzias1965}.
Predictions of neutrino number density anisotropies have been attempted, with an eye toward their
effect on the CMB power spectrum \cite{Trotta}, however there are too many free parameters for such
work to be conclusive. Instead, our approach in analogizing microwave background temperature
variations to the neutrino background seems to have been largely avoided, but there is no reason we
should expect primordial neutrinos not to undergo SW and ISW effects. The one notable exception is
Ref.~\cite{Hu:1995fq} in which the full Boltzmann equations for the neutrino brightness perturbation
is evolved in a cold dark matter dominated universe with $\Omega_m = 1$. Although this is a standard by-product of CMB
calculations, their Figure 12 is the only other display of the CNB anisotropy power spectrum of which
we are aware. 

It would be surprising if there are foreground neutrino sources which contaminate the CNB. Although
there exists a rich background of astrophysical, atmospherical, and terrestrial neutrino sources, it
is hard to believe that the spectrum extends down to $1.9$~K, considering that the exclusive
weak-force interaction would prevent any process but redshifting from reducing conventional neutrino
energies by that many orders of magnitude.

We also note that none of the current proposals for CNB detection would be capable of observing the
angularly dependent anisotropy behavior described in this paper. Of all prospects for measurement,
annihilation with UHE neutrinos due to a resonance with Z bosons would likely give the most
unequivocal proof for the existence the CNB \cite{Gelmini:2004hg}. In this scenario, a burst of
cosmic rays from a topological defect is anomalously absorbed by the CNB, causing observers of the
UHE$\nu$'s to see dips at characteristic energies corresponding to the three masses of the CNB
neutrinos. However the discovery of a topological defect required to produce the cosmic rays would be
so astounding in its own right that this method does not currently seem promising. 

We expect the CNB anisotropy is nearly impossible to detect directly. First, the cross section for a
relic neutrino to scatter off a nucleus is tremendously low. Estimating $\sigma \approx G_F^2
E_{\nu}^2 \simeq 4 \times 10^{-64} {\rm cm}^2$, this makes dark matter detection look easy. Second, a
threshold energy $E_\nu \sim 10^{-4}$~eV, many orders of magnitude below the detection sensitivity of
current experiments, would be needed to see the peak of the CNB spectrum; more still to see the
minute anisotropic temperature variations. Third, the large volume of material necessary to build a
suitable detector makes for poor angular resolution. Nonetheless, the anisotropic CNB is there.

The hurdles in even observing the CNB are so significant that to speculate as to its {\it angular}
appearance must seem somewhat presumptuous. We make no assertions about the feasibility of such a
prospect beyond noting that many of the accomplishments in neutrino physics experimentation were
completely unanticipated or considered unattainable until shortly before their implementation. We
recall Wolfgang Pauli's remark shortly after conceiving of the neutrino in 1930 \cite{Hoyle}:  ``I've
done a terrible thing today, something which no theoretical physicist should ever do. I have
suggested something that can never be verified experimentally." Hopefully, a discussion of CNB
properties will someday cease to seem as exclusively theoretical as it does today, just as the
neutrino itself once did to Pauli.

%%%%%%%%%%%%%%%%%%%%%%%%%%%%%%%%%%%%%%%%%%%%%%%%%%%%%%%%%%%%%%%%%%%%%%%%%%%%%%%%%%%%%%%%%%%%%
\begin{acknowledgments}
R.C. was supported in part by NSF AST-0349213 at Dartmouth. 

\end{acknowledgments}

\vfill

%%%%%%%%%%%%%%%%%%%%%%%%%%%%%%%%%%%%%%%%%%%%%%%%%%%%%%%%%%%%%%%%%%%%%%%%%%%%%%%%%%%%%%%%%%%%%

%%%%%%%%%%%%%%%%%%%%%%%%%%%%%%%%%%%%%%%%%%%%%%%%%%%%%%%%%%%%%%%%%%%%%%%%%%%%%%%%%%%%%%%%%%%%%

\begin{thebibliography}{99}

\bibitem{KolbTurner} 
  E.~W.~Kolb and M.~S.~Turner, The Early Universe, Addison-Wesley (1990).

\bibitem{Hannestad:2006zg}
  S.~Hannestad,
  %``Primordial neutrinos,''
  arXiv:hep-ph/0602058.

\bibitem{Kayser}
  B.~Kayser, ``Neutrino Mass, Mixing, and Flavor Change," in The Particle Data Group (S. Eidelman et al.), Phys. \ Lett. \ {\bf B592}, 1 (2004).

\bibitem{SKref} 
  S.~Fukuda {\it et al.}, [Super-Kamiokande Collaboration], Phys. Lett. {\bf B539}, 179 (2002). 
  
\bibitem{SNOref} 
  Q.~R.~Ahmad {\it et al.}, [SNO Collaboration], Phys. Rev. Lett. {\bf 89}, 011301 (2002);
  {\it ibid.}, 011302 (2002).
  
\bibitem{Kamref} 
  K.~Eguchi {\it et al.}, [KamLAND Collaboration], Phys. Rev. Lett. {\bf 90}, 021802 (2003).

\bibitem{Ma:1995ey}
  C.~P.~Ma and E.~Bertschinger,
  %``Cosmological perturbation theory in the synchronous and conformal Newtonian
  %gauges,''
  Astrophys.\ J.\  {\bf 455}, 7 (1995).

\bibitem{SachsWolfe}
  R.~K.~Sachs, A.~M.~Wolfe, Astrophys.\ J.\  {\bf 147}, 73 (1967).

\bibitem{Rees}
  M.~J.~Rees, D.~W.~Sciama, Nat. {\bf 217}, 511 (1968).
     
\bibitem{Mukhanov}
  V.~F.~Mukhanov, H.~A.~Feldman and R.~H.~Brandenberger,
  %``Theory Of Cosmological Perturbations. Part 1. Classical Perturbations. Part
  %2. Quantum Theory Of Perturbations. Part 3. Extensions,''
  Phys.\ Rept.\  {\bf 215}, 203 (1992).
  
\bibitem{Spergel:2006hy}
  D.~N.~Spergel {\it et al.},
   ``Wilkinson Microwave Anisotropy Probe (WMAP) three year results:
  %Implications for cosmology,''
  arXiv:astro-ph/0603449.
  
\bibitem{CMBfast}
  See CMBfast ({\tt www.cmbfast.org}) documentation for an explanation of 
  the WMAP normalization.

\bibitem{Seljak:1996is}
  U.~Seljak and M.~Zaldarriaga,
  %``A Line Of Sight Approach To Cosmic Microwave Background Anisotropies,''
  Astrophys.\ J.\  {\bf 469}, 437 (1996).
  
\bibitem{Hu:1995fq}
  W.~Hu, D.~Scott, N.~Sugiyama and M.~J.~White,
  % ``The Effect Of Physical Assumptions On The Calculation Of Microwave
  %Background Anisotropies,''
  Phys.\ Rev.\ D {\bf 52}, 5498 (1995).
   
\bibitem{DegenerateNeutrinos}
  K.~Ichikawa and M.~Kawasaki,
  %``Remarks on the cosmic density of degenerate neutrinos,''
  Phys.\ Rev.\ D\ {\bf 67}, 063510 (2003).

\bibitem{Khlopov1981}
  Ya.~B.~Zeldovich and M.~Khlopov,
  %``The neutrino mass in elementary particle physics and in big bang cosmology,''
  Sov. Phys. Usp. {\bf 24}, 755 (1981).
  
\bibitem{Lesgourgues:2006nd}
  J.~Lesgourgues and S.~Pastor,
  %``Massive Neutrinos And Cosmology,''
  Phys.\ Rept.\  {\bf 429}, 307 (2006).

\bibitem{Dicke1965}
  R.~H.~Dicke, P.~J.~E.~Peebles, P.~G.~Roll, and D.~T.~Wilkinson,
  Astrophys.\ J.\  {\bf 142}, 414 (1965).
  
\bibitem{Gerstein1966}
  S.~S.~Gerstein and Ya.~B.~Zeldovich,
  % ``Rest mass of muonic neutrino and cosmology,''
  Sov. Phys. JETP Lett. {\bf 4}, 120 (1966).

\bibitem{Penzias1965}
  A.~A.~Penzias and R.~W.~Wilson,
  Astrophys.\ J.\  {\bf 142}, 419 (1965).
     
\bibitem{Trotta}
R.~Trotta and A.~Melchiorri,  Phys.\ Rev.\ Lett.\  {\bf 95}, 011305 (2005).

\bibitem{Gelmini:2004hg}
  G.~B.~Gelmini,
  %``Prospect For Relic Neutrino Searches,''
  Phys.\ Scripta {\bf T121}, 131 (2005).
  
\bibitem{Hoyle}
F.~Hoyle, Proc. Roy. Soc. London A {\bf 301}, 171 (1967).
 %This is the reference given in astro-ph/0603603. I think my access may have been withdrawn and I am having trouble accessing it.
    
\end{thebibliography}
\end{document}